\documentclass[12pt]{article}\usepackage[hyperfootnotes=false]{hyperref}
   \usepackage{epsfig}
      \usepackage{amsmath}
  \usepackage{graphicx}
  \newlength{\abstractwidth}
  \renewcommand{\thefootnote}{\fnsymbol{footnote}}
  \renewcommand{\thanks}[1]{\footnote{#1}} 
  \newcommand{\starttext}{
  \setcounter{footnote}{0}
  \renewcommand{\thefootnote}{\arabic{footnote}}}
  \renewcommand{\theequation}{\thesection.\arabic{equation}}
  \newcommand{\be}{\begin{equation}}
  \newcommand{\bea}{\begin{eqnarray}}
  \newcommand{\eea}{\end{eqnarray}}
  \newcommand{\beq}{\begin{equation}}
  \newcommand{\ee}{\end{equation}}
  \newcommand{\eeq}{\end{equation}}

  \def\ba{\begin{eqnarray}}
  \def\ea{\end{eqnarray}}

  \def\12{{1 \over 2}}

  \def\d{\partial}

  \def\cc{cosmological constant }
  
  \def\simleq{\; \raise0.3ex\hbox{$<$\kern-0.75em
      \raise-1.1ex\hbox{$\sim$}}\; }
   \def\simgeq{\; \raise0.3ex\hbox{$>$\kern-0.75em
      \raise-1.1ex\hbox{$\sim$}}\; }

\def\ba{\bf{a}}

  \def\h3{{\cal{H}}_3}

\def\o3{\Omega_3}

\def\O2{\Omega_2}
\def\o{\omega}
 \def\d2{$dS_{2+1}$}
\def\21{$(2+1)$-dimensional}
\def\31{$(3+1)$-dimensional}
 \def\bi{\begin{itemize}}
  \def\ei{\end{itemize}}

\begin{document}
  \renewcommand{\theequation}{\thesection.\arabic{equation}}


  \begin{titlepage}
  \rightline{}
  \bigskip

  \bigskip\bigskip\bigskip\bigskip

    \centerline{\Large \bf {Was There a Beginning?}}
    \bigskip

  \bigskip \bigskip

  \bigskip\bigskip
  \bigskip\bigskip

  \begin{center}
  {{ Leonard Susskind}}
  \bigskip

\bigskip
Stanford Institute for Theoretical Physics and  Department of Physics, Stanford University\\
Stanford, CA 94305-4060, USA \\

\vspace{2cm}
  \end{center}

 \bigskip\bigskip
  \begin{abstract}


In this note I respond to Mithani and Vilenkin's claim that there must have been a beginning.

 \medskip
  \noindent
  \end{abstract}

  \end{titlepage}
  \starttext \baselineskip=17.63pt \setcounter{footnote}{0}


Mithani and Vilenkin have argued that the universe must have had a beginning \cite{Mithani:2012ii}. I will argue the opposite  point of view; namely, for all practical purposes, the universe was past-eternal.

To make the point simply, imagine Hilbertville, a one-dimensional semi-infinite city, whose border is at $x=0.$ The population is infinite and uniformly fills the positive axis  $x>0.$  Each citizen has an identical telescope with a finite power. Each wants to know if there is a boundary to the city. It is obvious that only a finite number of citizens can see the boundary at $x=0$. For the infinite majority the city might just as well extend to the infinite negative axis. Thus, assuming he is typical, a citizen who has not yet studied the situation should bet with great confidence that he cannot detect a boundary. This conclusion is independent of the power of the telescopes as long as it is finite.

Now let us consider some of the examples in Mithani and Vilenkin's note. First de Sitter space. For generality we can consider a landscape of positive \cc \ vacua with a bounded vacuum energy greater than zero. We may assume an initial starting point, for example an initial vacuum on the landscape. As explained in \cite{Harlow:2011az} and \cite{Susskind:2012pp} if we follow a causal patch it will pass through an infinite number of Boltzmann fluctuations that will sample all vacua. Most observer's will be freaks but even if we condition on normal observers, there will be an infinite number of them. Given any finite time $T,$  all but a finite number of them will occur later than that time. By later I mean after the initial starting point. Thus if they think they are typical, the citizens should bet---again with overwhelming confidence---that they cannot detect a beginning.

Next consider the so-called cyclic universe---eternal to the past and to the future. Obviously, if such a thing could exist it would have no beginning, but it cannot exist; not without violating the second law of thermodynamics.

\begin{figure}[h]
\begin{center}
\includegraphics[scale=.4]{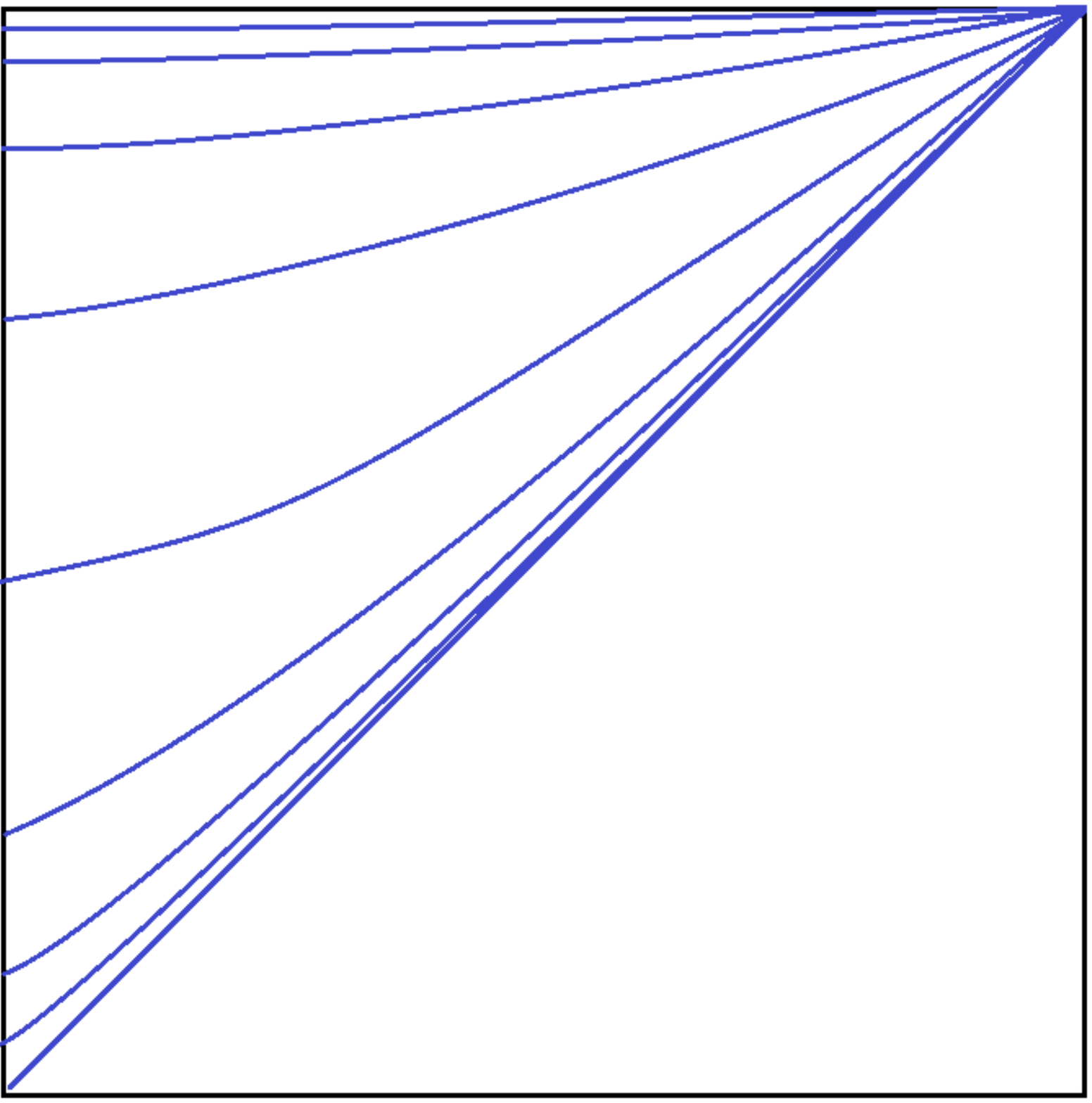}
\caption{Penrose Diagram of an expanding cyclic universe. The blue curved lines represent bounces.}
\label{f1}
\end{center}
\end{figure}

To avoid a heat-death the proponents of a cyclic universe assume that with each cycle the universe expands by a linear factor $L,$ thereby diluting the entropy \cite{Steinhardt:2002ih}. Thus the cyclic universe expands exponentially. Averaging over a cycle, the cyclic universe is just de Sitter space in flat-slicing. This is show in Figure \ref{f1}.

\begin{figure}[h]
\begin{center}
\includegraphics[scale=.4]{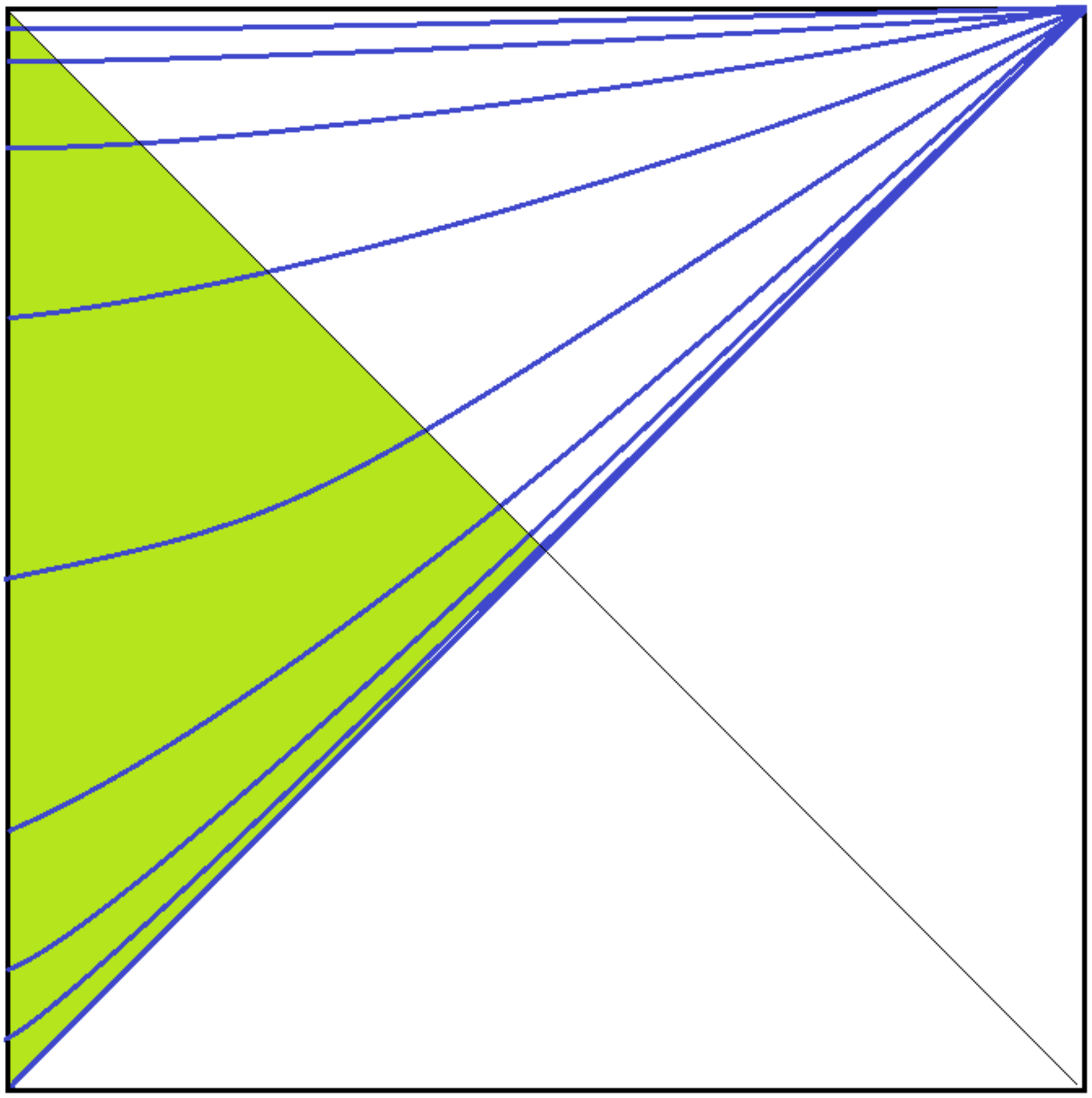}
\caption{Causal patch in a cyclic universe. The green area is the region covered by the analog of static coordinates in de Sitter space. The observer sees  perpetual motion.}
\label{f2}
\end{center}
\end{figure}

Now consider an observer at $r=0$. Such an observer is obviously surrounded by an event-horizon as in Figure \ref{f2}. From his point of view the universe is a bounded region with a finite entropy bound. In the causal patch the entropy does not dilute.
It is evidently a perpetual motion machine that violates the second law. Thermodynamically it must run down in a finite time. Thereafter it will just be the causal patch of de Sitter space. Periodicity must give way to the recurrent behavior of the previous case.

Finally let us consider eternal inflation with terminal vacua. Although nothing I say will depend on it, we can visualize the multiverse using the tree-model of \cite{Harlow:2011az}. There is a paradox in this case which I explained in \cite{Susskind:2012pp}. We assume a landscape with $e^N$ de Sitter vacua.
Suppose we  follow a causal patch from the root of the tree. It starts with vacuum type $n$ and after about $N$ transitions it will enter a terminal vacuum. The probability for surviving more than $N$ transitions goes to zero. Thus an observer should bet that he is within $N$ transitions of the root. With a sufficiently powerful telescope he should be able to see the beginning.

On the other hand consider the global view. Even though any given causal patch dies in a terminal, the number of causal patches increases exponentially with time. Therefore the vast majority of observers are very high up on the tree. At late times the statistics of the  vacuum-types is governed by an attractor called a fractal flow \cite{Harlow:2011az} \cite{Susskind:2012pp}  which is completely insensitive to the initial condition. If a multiversal citizen knows these things he will bet that he is too late to detect any evidence of the root.

Combing the Mithani-Vilenkin's observations \cite{Mithani:2012ii} with the ones in this note, we may conclude that there is a beginning, but
 in any kind of inflating cosmology the odds strongly (infinitely) favor the beginning to be so far in the past that it is effectively at minus infinity. More precisely, given any $T$ the probability is unity that the beginning was more than $T$  time-units ago.

  \end{document}